\documentclass[final,5pt,twocolumn,amssymb,prd]{revtex4-1}
\usepackage{pgfplots}
\usepackage{latexsym,epsfig}
\usepackage{graphicx}
\usepackage{ulem}
\usepackage{dcolumn}
\usepackage{color}
\usepackage{amsthm,amsmath}

\begin{document}

\title{Reply to Comment on ``New physics constraints from atomic parity violation in $^{133}$Cs''}

\author{B. K. Sahoo$^a$\footnote{Email: bijaya@prl.res.in},  B. P. Das$^{b,c}$ and H. Spiesberger$^d$}

\affiliation{$^a$Atomic, Molecular and Optical Physics Division, Physical Research Laboratory, Navrangpura, Ahmedabad-380009, India\\
$^b$Department of Physics, School of Science, Tokyo Institute of Technology, 2-1-2-1-H86 Ookayama Meguro-ku, Tokyo 152-8550, Japan \\
$^c$Centre for Quantum Engineering Research and Education, TCG Centres for Research in Science and Technology, Sector V, Salt Lake, Kolkata 70091, 
India\\ 
$^d$PRISMA$^+$ Cluster of Excellence, Institut f\"{u}r Physik, Johannes Gutenberg-Universit\"{a}t, D-55099 Mainz, Germany}

\begin{abstract}
In Phys. Rev. D {\bf 103}, L111303 (2021), we had reported an improved calculation of the nuclear spin-independent parity violating electric dipole 
transition amplitude ($E1_{PV}$) for the $6s ~ ^2S_{1/2} - 7s ~ ^2S_{1/2}$ transition in $^{133}$Cs by employing a relativistic coupled-cluster (RCC) 
theory. 
%The combination of the most accurate (0.3\%) measurement with our calculation enabled us to determine a new value for the nuclear weak charge 
%$Q_W=-73.71(26)_{ex} (23)_{th}$. The Standard Model (SM) prediction of this quantity is $Q_W^{\text{SM}}=-73.23(1)$. 
In a recent Comment, B. M. Roberts and J. S. M. Ginges have raised questions about our calculation of the so-called Core contribution
to $E1_{PV}$. Our result for this contribution does not agree with theirs, but is in agreement with results from previous calculations where 
this contribution is given explicitly. In our reply, we explain in detail the validity of the evaluation of our core contribution. We emphasize
that the Main, Core and Tail contributions have been treated on an equal footing in our work unlike the sum-over-states calculations. We also 
address their concerns about our approximate treatment of the contributions from the QED corrections, which was not the aim of our work, but was 
carried out for completeness. Nonetheless, conclusion of our above mentioned paper is not going to affect if we replace our estimated QED 
contribution to $E1_{PV}$ by earlier estimation.
\end{abstract}

\date{Received date; Accepted date}

\maketitle

\section{Background}

In order to set the scene for our reply to the Comment by B. M. Roberts and J. S. M. Ginges \cite{Ginges} on our paper \cite{Sahoo}, we would 
like to mention that the calculation of the parity violating electric dipole transition amplitude of $^{133}$Cs has a long history. In particular,
there was an unsettled issue about the contributions from the occupied orbitals, referred as ``Core'' contribution, to the
parity violating electric dipole amplitude ($E1_{PV}$) of the $6s ~ ^2S_{1/2} \rightarrow 7s ~ ^2S_{1/2}$ transition in the $^{133}$Cs atom 
from the Dirac-Coulomb (DC) Hamiltonian that differed by about 200\% between the previous two high-precision calculations reported in Refs. 
\cite{Porsev} and \cite{Dzuba}. in p.~26 of the review article by M. S. Safronova, D. Budker, D. DeMille, D. F. J. 
Kimball, A. Derevianko and C. W. Clark \cite{Safronova}, it is explicitly mentioned that:
``{\it One of us, A.D., thinks that the correction to the contribution of highly excited states (Dzuba et al., 2012) may have come from
the use of many-body intermediate states by Dzuba et al. (2012) that is inconsistent with the one employed by Porsev, Beloy, and Derevianko 
(2009), as the summation over intermediate states while evaluating $k_{PV}$ must be carried out over a complete set and thereby the results of 
Dzuba et al. (2012) require revision. This matter remains unresolved at present but new methods are being developed to address it. The 
ever-increasing power of computation is anticipated to bring further improvements in the atomic-structure analysis.}''.

Similar comments were also made in another unpublished article \cite{Derevianko} soon after the above review. Our work in Ref.~\cite{Sahoo} 
was mainly devoted to addressing the above issue by directly solving the first-order perturbed wave functions due to the weak interaction with
reference to the zeroth-order wave functions of the DC Hamiltonian using the relativistic coupled-cluster (RCC) theory. This overcomes the 
shortcomings of the sum-over-states RCC theory approach of Ref.~\cite{Porsev} to include the Core contributions to all-orders in the singles and 
doubles RCC theory approximation (RCCSD method) and singles, doubles and triples RCC theory approximation (RCCSDT method). In view of the
practical  limitations from a computational point of view, triple excitations were considered only for selected low-lying orbitals. 
In addition, the corrections due to the exchange of transverse photons between pairs of electrons, known as the Breit interaction, and the quantum 
electrodynamics (QED) interactions were included to improve the  atomic wave function in the RCCSDT method. This method was evaluated to 
use the Main, Core and Tail contributions on an equal footing. Moreover, the Hamiltonians representing electromagnetic ($H_{em}$) and weak 
($H_{weak}$) interactions were treated in a consistent manner. The accuracies of these calculations were estimated by comparing the results of 
calculated spectroscopic properties with those available from experiments as has 
been done traditionally, e.g. in Ref.~\cite{Porsev}. In fact, well before our results were published in Ref.~\cite{Sahoo} they were uploaded on 
arXiv in order to seek responses from others \cite{Sahoo1,Sahoo2}. Nonetheless, we are convinced that the arguments presented by Roberts and 
Ginges leave a lot to be desired and they do not change the conclusions we arrived at in Ref.~\cite{Sahoo}. 

Our arguments are the following:

(i) In Refs.~\cite{Porsev,Dzuba}, the Breit and QED corrections to $E1_{PV}$ and other properties were borrowed from previous calculations, 
which were reported using lower-order many-body methods and basis functions different from those used in the DC calculations. In our work, the 
above interactions were included along with the DC Hamiltonian in $H_{em}$ for the determination of atomic wave functions in Ref.~\cite{Sahoo}. 
This enabled us to show that the Breit and QED contributions from the RCC theory were different from the earlier values using less rigorous 
many-body methods. Using a single rigorous relativistic many-body method in which $H_{em}$ included the DC, Breit and QED interactions serves
as a more realistic way of estimating uncertainties in the calculated atomic properties rather than either borrowing corrections from other
calculations or by scaling the atomic wave functions to produce the results matching with the experimental values even though small corrections 
from the QED effects to the property evaluation were neglected. The important point is that the QED effects were included at the same level of 
approximations as those of the DC and Breit interactions, and the estimated QED contributions matched reasonably well for all the properties with 
the earlier evaluations. Similar to Ref. \cite{Porsev}, the accuracies of the final results of various properties were analyzed by comparing with
those of their experimental values in Ref. \cite{Sahoo}. 
 
\begin{table}[t]
\caption{Comparison of contributions from the `Core' and `Virtual' orbitals to the $E1_{PV}$ amplitude (in $-i (Q_{W}/N) ea_0 \times 10^{-11}$)
of the $6s ~ ^2S_{1/2} - 7s ~ ^2S_{1/2}$ transition in $^{133}$Cs using the Dirac-Coulomb Hamiltonian from various works. We have also mentioned 
the many-body methods and approaches employed in determination of these contributions. The signs of the Core orbital contributions from 
Refs.~\cite{Ginges,Dzuba}, marked in bold fonts, differ from the other works.}
 \begin{ruledtabular}
  \begin{tabular}{l l c c c }
  Method    & Approach & Core & Virtual & Reference \\
 \hline 
 HF  & {\it ab initio}  &  $-0.00174$  &   & \cite{Ginges}  \\
 RPA  & {\bf ab initio}  &  {\bf 0.00170}  &   & \cite{Ginges}  \\
   RPA  & {\bf scaled}  &  {\bf 0.00259}  &   & \cite{Ginges}  \\
  BO$+$RPA & {\bf  ab initio} & {\bf 0.00181} &  & \cite{Ginges} \\
 BO$+$RPA & {\bf scaled} & {\bf 0.00181} &  & \cite{Ginges} \\
 \hline 
 HF & {\it ab initio} & $-0.0017$  & 0.7401 &  \cite{Sahoo} \\
 RCCSD & {\it ab initio} & $-0.0019$  & 0.9006 &  \cite{Sahoo} \\
 RCCSDT & {\it ab initio}  & $-0.0018$ & 0.9011 &  \cite{Sahoo}  \\
 \hline 
 \multicolumn{2}{l}{Lower-order} & $-0.0020$ & & \cite{Porsev} \\
 RCCSDT & {\it sum-over}  &  & 0.9073 & \cite{Porsev} \\
 RCCSDT & {\it sum-over$+$scaled} &  &  0.9018 & \cite{Porsev} \\
 \hline 
  HF  & {\it ab initio}  &  $-0.00174$  &   & \cite{Dzuba}  \\
  RPA  & {\bf scaled}  &  {\bf 0.00259}  &   & \cite{Dzuba}  \\
 BO$+$RPA & {\bf ab initio} & {\bf 0.00170} & 0.8949 & \cite{Dzuba} \\
 BO$+$RPA & {\bf scaled} & {\bf 0.00182} & 0.8920  & \cite{Dzuba} \\
 \hline 
 \multicolumn{5}{c}{\underline{Earlier reported Core contributions}} \\
  RCCSD & {\it ab initio} & $-0.002$   &  & \cite{Das} \\
  RCCSD & {\it ab initio} & $-0.002$   &  & \cite{Sahoo4} \\
  RCCSD & {\it ab initio} & $-0.0019$   & & \cite{Sahoo5} \\
  \multicolumn{2}{l}{Lower-order} & $-0.002(2)$ & & \cite{Blundell} \\
  \end{tabular}
 \end{ruledtabular}
 \label{tab1}
\end{table}
 
(ii) The Core contribution to $E1_{PV}$ (given in units of $-i (Q_{W}/N) ea_0 \times 10^{-11}$ here onwards with nuclear weak charge $Q_W$ and 
neutron number $N$) of the $6s ~ ^2S_{1/2} \rightarrow  7s ~ ^2S_{1/2}$ transition in $^{133}$Cs were evaluated using lower-order methods as 
$-0.0020$ while the Hartree-Fock (HF) value is about  $-0.0017$ \cite{Ginges,Sahoo,Dzuba}. This Core contribution was reported to be about
$-0.0019$ using the RCCSD method~\cite{Das,Sahoo4,Sahoo5} by us much before $E1_{PV}$ calculations were reported in Refs.~\cite{Ginges,Porsev,
Dzuba}. In Ref.~\cite{Sahoo}, it was investigated using the RCCSDT method to find whether higher-order effects are responsible for the large 
difference in the results between Refs.~\cite{Porsev} and \cite{Dzuba}. It was, however, found that the result changed marginally to $-0.0018$. 
Our RCCSD method was also employed earlier to estimate the Core contributions to $E1_{PV}$ of Ba$^+$ \cite{Sahoo3} and Ra$^+$ \cite{Wansbeek}, 
and Roberts and Ginges have acknowledged saying that they have reproduced our result for Ra$^+$. The reasons for the disagreement for the Core
contribution for Cs between the results referred to by Roberts and Ginges and ours is the difference in the physical effects included in the 
two works and the improper scaling of the atomic wave functions to compensate for the missing physical effects in the evaluation of $E1_{PV}$. Since 
a $S \leftrightarrow S$ transition is involved in the Cs atom, there are huge cancellations in the contributions to $E1_{PV}$ from both the 
states. In contrast, contributions to $E1_{PV}$ arise mainly from the $S$ state of a  $S \leftrightarrow D_{3/2}$ transition in Ra$^+$. 
   
%\section{$E1_{PV}$ calculations and their limitations}
%\subsection{$E1_{PV}$ evaluation procedure}

\section{$E1_{PV}$ evaluation procedure}

Using the sum-over-states approach, $E1_{PV}$ is evaluated as \cite{Porsev,Blundell}
\begin{eqnarray}
 E1_{PV} &\simeq & \frac{\lambda}{\cal N} \sum_{I \ne f} \frac{\langle \Psi_f^{(0)} | H_{weak} | \Psi_I^{(0)} \rangle \langle \Psi_I^{(0)} | D | \Psi_i^{(0)} \rangle}
 {(E_f^{(0)} - E_I^{(0)}) } \nonumber \\
 && + \frac{\lambda}{\cal N} \sum_{I \ne i} \frac{\langle \Psi_f^{(0)} | D | \Psi_I^{(0)} \rangle \langle \Psi_I^{(0)} | H_{weak} | \Psi_i^{(0)} \rangle}
 {(E_i^{(0)} - E_I^{(0)})} ,
\label{e1pnc1}
\end{eqnarray}
where $\lambda=\frac{G_F}{2\sqrt{2}}Q_{W}$, $| \Psi_{k=i,f,I}^{(0)} \rangle$ are the wave functions corresponding to $H_{em}$ with energies 
$E_k^{(0)}$, $D$ is the electric dipole (E1) operator, $I$ denotes all the intermediate states and ${\cal N}$ is the normalization constant. 
This approach has the limitation that contributions only from a few low-lying bound states can be evaluated through this approach (usually referred
as ``Main'' contribution), whereas contributions from the occupied orbitals (referred as ``Core'' contributions) and high-lying bound states 
including continuum (denoted as ``Tail'') contributions are challenging to treat accurately for practical reasons. Thus, the Core and Tail are 
usually estimated by applying either a lower-order theory or mixed many-body methods. The limitations of using a mixed approach to determine 
$E1_{PV}$ amplitude is that the correlations among the Core, Main and Tail parts (the latter two together referred as `Virtuals' from here onwards)
and corrections due to ${\cal N}$ do not lend themselves to being included naturally at the same level of approximation. In particular, the 
interplay of the correlations among all these contributions cannot be accounted for by the above mentioned many-body methods that have been used 
in the sum-over-states approach. In addition, the interplay of the correlations among the electromagnetic interactions and weak interactions 
are also not accounted for properly in such an approach. This clearly suggests that it is imperative to consider the Core, Main and Tail 
contributions and also the interplay of the correlations between the interactions due to $H_{em}$ and $H_{weak}$ on an equal footing, and 
estimate corrections due to ${\cal N}$ from all these contributions at the same level of approximation.

\begin{figure}[t]
%centering
\includegraphics[width=8.5cm,height=5cm]{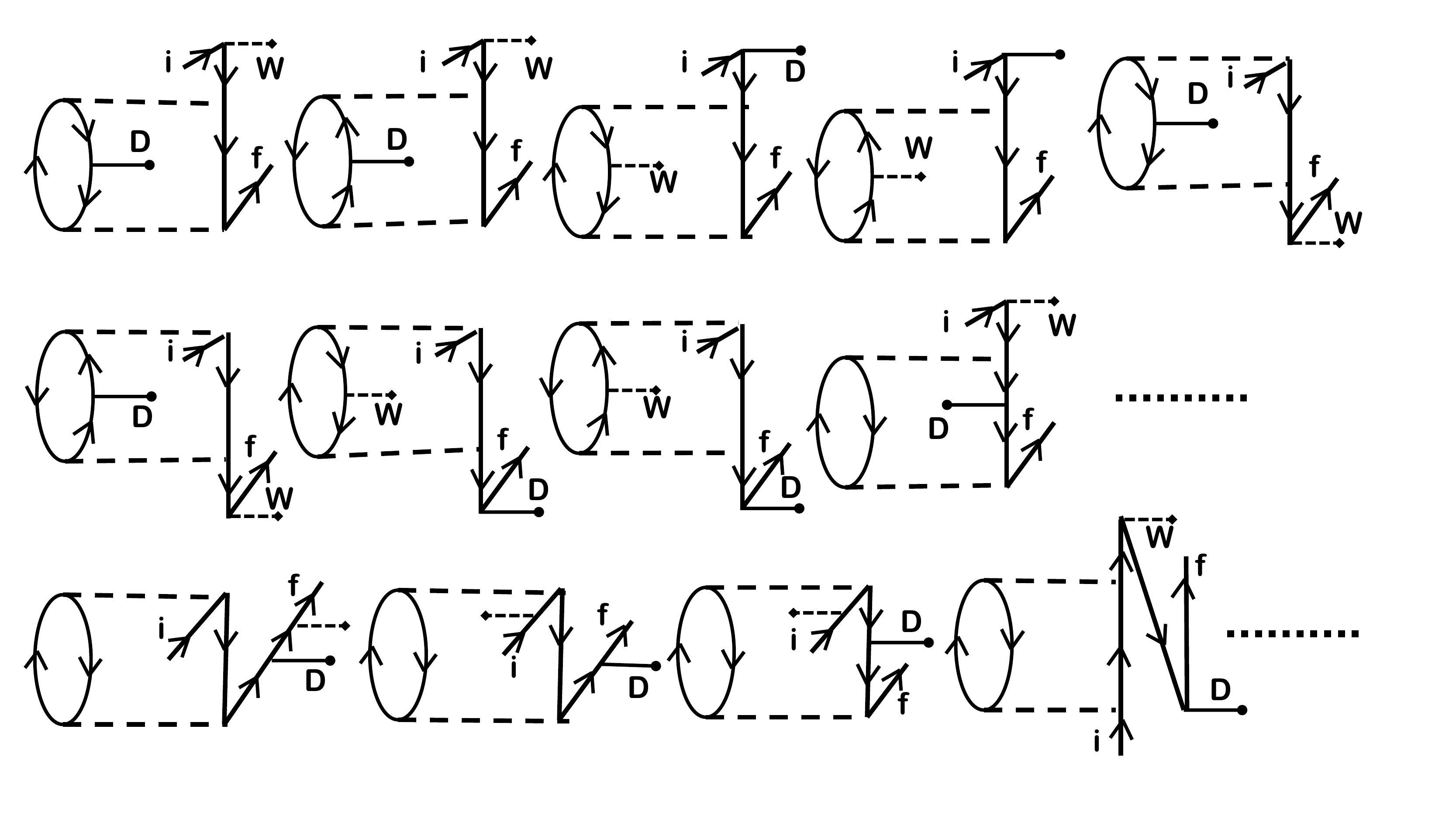}
\caption{\label{fig1} A few examples of Goldstone diagrams representing non-RPA effects to the Core contributions in the determination of 
$E1_{PV}$ amplitude in the one-valence atomic systems. Contributions from such diagrams to $E1_{PV}$ of the $6s ~ ^2S_{1/2} - 7s ~ ^2S_{1/2}$ 
transition in $^{133}$Cs cancel out strongly with that arise due to RPA. Here lines with upward arrows denote virtuals, lines with downward arrows
denote occupied orbitals, $i$ and $f$ are the initial and final valence orbitals respectively, lines with $D$ represent the E1 operator,  
lines with $W$ correspond to the $H_{weak}$ operator and the dashed lines stand for $H_{em}$ operator.} 
\end{figure}

The shortcomings of the sum-over-states approach are circumvented in Ref. \cite{Sahoo} by calculating $E1_{PV}$ as
\begin{eqnarray}
  E1_{PV} & \simeq & \lambda \frac{\langle \Psi_f^{(1)} | D | \Psi_i^{(0)} \rangle + \langle \Psi_f^{(0)} | D | \Psi_i^{(1)} \rangle}
  {\sqrt{\langle \Psi_f^{(0)} | \Psi_f^{(0)} \rangle \langle \Psi_i^{(0)} | \Psi_i^{(0)} \rangle}}  \nonumber \\
   &=& \frac{\lambda}{{\cal N}} \left [ \langle \Psi_f^{(1)} | D | \Psi_i^{(0)} \rangle + \langle \Psi_f^{(0)} | D | \Psi_i^{(1)} \rangle \right ],
\label{e1pnc}
\end{eqnarray}
where $| \Psi_{k=i,f}^{(1)} \rangle$ denotes the first-order perturbed wave functions due to $H_{weak}$, respectively, which are obtained by 
solving the following inhomogeneous equations in the RCC theory framework
\begin{eqnarray}
 (H_{em} -E_k^{(0)})  | \Psi_k^{(1)} \rangle = - H_{weak} | \Psi_k^{(0)} \rangle .
\end{eqnarray}
Since both $| \Psi_k^{(0)} \rangle$ and $| \Psi_k^{(1)} \rangle$ wave functions are obtained using the same RCC theory, contributions from the 
Core and Virtual orbitals are embedded in the evaluation of Eq. (\ref{e1pnc}). In Ref. \cite{Sahoo}, we had made a special effort to present 
results in terms of the Core, Main and Tail contributions to facilitate understanding of these contributions to address the unsettled issues
raised in Refs. \cite{Safronova,Derevianko}.

\begin{table}[t]
\caption{The estimated contributions from the Breit interaction to the EAs (in cm$^{-1}$) in Cs reported in various works.}
\begin{ruledtabular}
\begin{tabular}{lccccc} 
  Method  & $6S$  & $6P_{1/2}$ & $7S$ & $7P_{1/2}$ & $8P_{1/2}$ \\   
 \hline \\ 
RMBPT(3) \cite{Derevianko1}   &   2.6   &  $-7.1$  &  0.26   & $-2.5$    &       \\
RMBPT \cite{Kozlov} & 4.39  & $-8.78$ & 0.0 & $-2.19$ & \\ 
SD \cite{Safronova1} & 1.1   &  $-6.9$ & $-0.72$  & $-2.6$ & \\ 
RCCSD \cite{Eliav}   & 1.0 & $-7.0$ & $0.0$& $-3.0$ & \\
RCCSDT \cite{Sahoo} & $-0.60$  & $-7.81$  & $-0.65$    & $-2.61$   &  $-1.21$  \\
\end{tabular}
\end{ruledtabular}
\label{tab2}
\end{table}

\section{Responses to Comments}

Roberts and Ginges start their Comment by referring to the estimated corrections due to the QED interactions in Ref. \cite{Sahoo}. As we have 
mentioned earlier, there have not been significant differences between the estimated QED corrections to $E1_{PV}$ in Cs, though they vary 
slightly between the different calculations, the major disagreement between Refs. \cite{Porsev} and 
\cite{Dzuba} was on the Core contribution to $E1_{PV}$ where there was a sign difference. The difference in the Tail contributions among these 
works was also quite large. Since it is not possible to consider the covariant form of $H_{em}$ for an atomic system for the determination of atomic 
wave functions, the dominant DC Hamiltonian is usually considered as its first approximation When necessary, contributions from the the Breit and QED 
interactions are estimated as corrections. Thus, it is imperative to include contributions from the DC Hamiltonian as accurately as possible. However, 
sometimes a less accurate many-body method is employed to include its contributions, and the calculated properties are rescaled to obtain the final 
results. Such a procedure cannot always give reliable results. In Ref. \cite{Sahoo}, we considered the Breit interaction potential and a model QED 
potential along with the DC Hamiltonian to include its contributions to all-orders in the residual interaction using the RCC theory. The authors of
Ref. \cite{Dzuba} used the Main contribution from Ref. \cite{Porsev} in their final result, but they had attempted to improve the Core and Tail 
contributions using the BO$+$RPA mixed many-body method. Since we evaluated $E1_{PV}$ of Cs in Ref. \cite{Sahoo} from first principles treating the
Main, Core and Tail contributions on an equal footing using the RCC theory, all the above contributions were embedded and they were inter-related. It
is evident from this that the results in Ref. \cite{Sahoo} were obtained in a more natural manner than those compared to other calculations. It does 
not extrapolate the results by scaling wave functions or using experimental data.

\begin{table}[t]
\caption{The reported Breit interaction contributions to the $A_{hyf}$ values (in MHz) in $^{133}$Cs from different methods.}
\begin{ruledtabular}
\begin{tabular}{lccccc} 
  Method  & $6S$  & $6P_{1/2}$ & $7S$ & $7P_{1/2}$ & $8P_{1/2}$ \\   
 \hline \\ 
RMBPT(3) \cite{Derevianko}  &  4.87  &  $-0.52$  &  1.15   & $-0.15$  &   \\
RMBPT \cite{Kozlov} & 5.0  & $-0.2$  & 0.8 & $0.0$ & \\ 
SD \cite{Safronova1} & $-4.64$  & $-0.87$  & $-0.83$ & $-0.29$ & \\
Analytic \cite{Sushkov} & 4.6  &   &  1.09  &  &  \\
RPA \cite{Blundell1} & 0.0 & $-1.25$ & $-0.05$ & $-0.39$ & \\ 
RCCSDT \cite{Sahoo} & 4.65     & $-0.18$  &  0.83    & $-0.04$ &  $-0.02$  \\
\end{tabular}
\end{ruledtabular}
\label{tab3}
\end{table}

To understand the possible reasons for the difference in the Core contribution to $E1_{PV}$ of Cs reported in Ref. \cite{Dzuba} from our RCC  
calculations \cite{Sahoo}, we present the Core and Virtual contributions to $E1_{PV}$ of Cs
from Refs. \cite{Ginges,Sahoo,Porsev,Dzuba} in Table \ref{tab1}. As can be seen from this table the HF value to the Core contribution is about 
$-0.0017$. This is different than what is being considered in Ref. \cite{Porsev}, so we do not know the basis for Roberts and Ginges 
referring to it as the HF value from Ref. \cite{Porsev}. Nonetheless, it is possible to find a one to one correspondence between the RPA terms and 
certain terms of the many-body perturbation theory (MBPT) as have been shown in Refs. \cite{Geetha,Geetha1}. Since RCC theory is an all-order 
perturbation theory, it contains all the RPA contributions and also takes into account the interplay between the RPA and the non-RPA effects. The 
RPA, however, misses out many contributions that appear in the RCC theory. In Fig. \ref{fig1}, we show a few selected lowest-order non-RPA 
correlation effects from the second-order MBPT method using the Goldstone diagram representation that are present to all-orders in the 
determination of Core contributions to $E1_{PV}$ using the RCC theory. We find a huge cancellations among the contributions from the non-RPA and 
the RPA correlation effects. The final small Core contribution to $E1_{PV}$ in Cs is the result of these strong cancellations. A perusal of the 
final result for $E1_{PV}$, listed in Table \ref{tab1}, from Refs. \cite{Ginges,Dzuba} suggests that it uses the scaled wave functions from the 
combined BO$+$RPA methods to obtain the Core contribution, whereas the ``Main'' result is taken from the RCC calculations of Ref. \cite{Porsev}.
It can also be noticed large difference between the Main contribution from the RCC theory with the combined BO$+$RPA calculations. However, only 
a small difference between the {\it ab initio} results for the Core contributions from RPA and the combined BO$+$RPA methods is noticed.
It suggests that perhaps the strong cancellation seen for the Core contribution from the RPA and non-RPA effects in the RCC theory is not 
happening through the scaled BO$+$RPA approaches. Also, we find that the scaling of contributions from the Virtuals, which also contains 
both RPA and non-RPA contributions and are the dominant ones, change by about 0.3\% while the scaled RPA contributions to the Core contributions 
are almost 34\%. So we have doubts over the efficacy of these scaling approaches, which are meant to capture the missing physical effects in the 
determination of wave functions using the BO$+$RPA methods. In fact, in another follow up work by Roberts {\it et al.} \cite{Roberts} mention that
they had missed out double-core polarization effects in Ref. \cite{Dzuba}, which appear naturally in the RCC theory. In addition, we find 
that there is about 1\% difference in the Virtual contributions between Refs. \cite{Porsev} and \cite{Dzuba}. This corroborates the point we had 
made above that the RPA and non-RPA contributions using the hybrid BO$+$RPA methods are inadequate.

\begin{table}[t]
\caption{The estimated QED interaction contributions to the EAs (in cm$^{-1}$) in the previous works.}
\begin{ruledtabular}
\begin{tabular}{lccccc} 
 Reference  & $6S$  & $6P_{1/2}$ & $7S$ & $7P_{1/2}$ & $8P_{1/2}$ \\   
 \hline \\ 
 \cite{Flambaum} & $-17.6$ & 0.4  & $-4.1$ & 0.1   & 0.05 \\ 
 \cite{Roberts2}   & $-0.069$\% & 0.006\% & $-0.040$\% & 0.004\% & 0.003\% \\
 \cite{Ginges1}   & $-25.28$  & 1.18 &  &  &  \\
 \cite{Sahoo} & $-20.53$ & 1.31    & $-5.09$    &  0.57     &   0.71    \\
\end{tabular}
\end{ruledtabular}
\label{tab5}
\end{table}

In Table \ref{tab2}, we present Breit contributions to EAs of the low-lying states of Cs from the previous works including from our RCCSDT method. In Ref. \cite{Porsev}, 
these contributions were borrowed from Ref. \cite{Derevianko1} that had employed the third-order relativistic many-body perturbation theory
(RMBPT(3)). As can be seen from this table, there are substantial differences among these results and they are mainly due to the electron 
correlation effects. Similarly, we also give the Breit contributions to the magnetic dipole hyperfine structure constants ($A_{hfs}$) in 
Table \ref{tab3} from different works and observe striking differences in the results. 
From Ref. \cite{Sahoo1}, one can observe how the Breit contributions to EAs and $A_{hf}$ vary from lower- to higher-order methods and
these variations are different for both the properties. This clearly indicates that the same scaling of wave functions may not give the correct 
results for properties described by operators with different ranks and radial behaviour (see also Ref. \cite{Sahoo6}). 

\begin{table}[t]
\caption{The reported QED interaction contributions to the $A_{hyf}$ values (in MHz) from the earlier calculations.}
\begin{ruledtabular}
\begin{tabular}{lccccc} 
  Reference  & $6S$  & $6P_{1/2}$ & $7S$ & $7P_{1/2}$ & $8P_{1/2}$ \\   
 \hline \\ 
 \cite{Cheng1,Cheng2} & $-9.7$  & $-0.05$ & $-2.30$  & $-0.02$ &  \\ 
 \cite{Ginges2} & $-8.8(1.5)$ &                       &   &  & \\
 \cite{Sahoo}  & $-7.28$  & 0.05     & $-1.51$  & 0.01    &  $\sim0.0$ \\
\end{tabular}
\end{ruledtabular}
\label{tab6}
\end{table}

It is desirable to have an exact expression to account for the QED effects for the determination of the atomic wave functions. However, there is no 
well-defined approach for incorporating the QED interactions to high-precision in many-electron systems and they are mostly accounted for using model 
potentials. To the best of our knowledge, the QED model potential defined in Refs. \cite{Voltoka,Shabaev}, which include contributions from the local
and non-local potentials to represent the self-energy (SE) interactions, are fairly reliable for estimating SE effects. In Ref. \cite{Sahoo}, we had 
used the expression for the SE interaction model potential defined in Ref. \cite{Ginges1,Flambaum} and the vacuum polarization effects were included 
in a manner similar to that used in Refs. \cite{Shabaev,Flambaum}. As provided by Ref. \cite{Voltoka}, the expression for estimating the SE 
contributions to the $A_{hf}$ constants can be divided into ``irreducible'' and ``reducible'' parts. The reducible part, which contains ultraviolet 
divergences, has two terms. Each of the terms of the reducible part may 
contribute significantly in the property evaluation, but the two contributions strongly cancel each other (e.g. see Ref. \cite{Voltoka}) resulting
in the largest contribution arises mainly from the irreducible part. The contribution of the QED model potential included in $H_{em}$ comes mainly 
from this irreducible part.

\begin{table}[t]
\caption{Comparison of contributions from the Breit and QED interactions to the $E1_{PV}$ amplitude (in $-i (Q_{W}/N) ea_0 \times 10^{-11}$) of 
the $6s ~ ^2S_{1/2} - 7s ~ ^2S_{1/2}$ transition in $^{133}$Cs from various methods employed in different works.}
 \begin{ruledtabular}
  \begin{tabular}{cccc}
 Breit & QED &  Method  &  Reference \\
 \hline \\
   $-0.0055(5)$ & $-0.0028(3)$  & RCCSDT   & \cite{Sahoo} \\
  & $-0.0029(3)$  & Correlation potential &   \cite{Flambaum}   \\
  $-0.0054$ &  & RMP(3)  &   \cite{Derevianko1} \\
  $-0.0045$ & $-0.27(3)$\%  & Local DHF potential &  \cite{Shabaev2} \\
 $-0.004$ &  & Optimal energy &    \cite{Kozlov} \\
  &   $-0.33(4)$\%   & Radiative potential &  \cite{Roberts1} \\
   $-0.0055$  &  & Correlation potential  &  \cite{Dzubab}\\
  \end{tabular}
 \end{ruledtabular}
 \label{tab8}
\end{table}

In Tables \ref{tab5} and \ref{tab6}, we present our estimated QED contributions to EA and $A_{hf}$ values, respectively, of the important low-lying 
states. Here, we have neglected the reducible contribution of the SE interaction to $A_{hf}$. As can be seen, there are large differences in the 
estimated QED contributions to EAs from the previously reported calculations. Thus, the electron correlation effects also change the magnitudes of the 
QED contributions. Therefore, it was imperative to give the estimated QED contribution from our RCC theory only rather than borrowing from another 
work in order to assess the quality of the calculated wave functions more reliably. This is why we had included the model QED potential through 
$H_{em}$ for the determination of atomic wave functions and the spectroscopic properties were evaluated using these wave functions. It was found that 
the estimated QED contribution to the ground state $A_{hf}$ value from this approach is $-7.28$ MHz, which is comparable to the more accurate calculations
$-9.7$ MHz \cite{Cheng1,Cheng2} and $-8.8(1.5)$ MHz \cite{Ginges2}. It can also be seen from Table \ref{tab8} that our estimated QED contribution 
to $E1_{PV}$ is $-0.0028(3)$, which is comparable to $-0.0024(3)$ used in Ref. \cite{Porsev} from Ref. \cite{Shabaev2} and $-0.0029(3)$ used 
in Ref. \cite{Dzuba} from Ref. \cite{Flambaum}. Here, we have quoted uncertainty only from the electron correlation effects by taking the 
difference in the results from the RCCSD and RCCSDT methods. Since our estimated QED contributions matches reasonably well with the previous estimations using 
other methods and they were treated in a consistent manner along with the DC and Breit interactions in Ref. \cite{Sahoo}, we did not focus on the 
missing small QED contributions to different properties. Since uncertainties for the many-body calculations cannot be estimated reliably, their
accuracies are analyzed by comparing the final calculated results with their experimental values. These comparisons are further used to estimate the 
uncertainty of $E1_{PV}$, which account for all possible missing physical effects including the higher level excitations and neglected QED effects. 
This is exactly what has been done in Ref. \cite{Sahoo} and is along the same lines as Ref. \cite{Porsev}. If Roberts and Ginges feel that this
is an issue then the accuracy of our calculated QED contribution to $E1_{PV}$ amplitude, can be determined by taking the QED correction either from 
Ref. \cite{Shabaev2} or Ref. \cite{Flambaum} as was done in Refs. \cite{Porsev,Dzuba}. In such a scenario, the final conclusion of our study would 
still remain the same. 
 
\section{Summary}

We have given possible reasons for the large differences between the core contributions to the parity violating electric dipole transition 
amplitude in Cs between Refs. \cite{Sahoo,Porsev} and \cite{Dzuba,Ginges}. This was the {\it prime motive} of carrying out our calculation of 
$E1_{PV}$ in Ref. \cite{Sahoo}. We have also explained why we have added both the Breit and QED interactions to the DC Hamiltonian in our 
calculations  rather than borrowing them from previous works. Since they are included in our calculations in a consistent manner, {\it albeit} 
the QED part being approximate, it would not be accurate to say that the uncertainty due to the QED effects in our result is larger than that in
Ref. \cite{Dzuba}. In fact, it is straight forward to use our results from the DC Hamiltonian and combine with the earlier estimated Breit and 
QED contributions to determine the uncertainty in the parity violating electric dipole transition amplitude of the Cs atom as has been done in 
Refs. \cite{Porsev,Dzuba}. In our view, that would not be appropriate owing to the fact that the many-body methods and basis functions used in 
those calculations would not be consistent. Even in such a case, our final result will not change significantly. However, it would be more 
important at this stage to include the neglected contributions from the Dirac-Coulomb Hamiltonian rigorously through the triple and quadruple 
excitations in the RCC theory in order to probe new physics beyond the Standard Model. Therefore, the concerns expressed by Roberts and Ginges 
in their Comment are misplaced.

\end{document}